\documentclass[12pt,a4paper]{elsarticle}
\usepackage[dvipsnames]{xcolor}
\usepackage{graphicx}
\usepackage{amssymb}
\usepackage{amsmath}
\usepackage{bm}
\usepackage{subfigure}
\usepackage[T1]{fontenc}
\usepackage{amsthm}
\usepackage{tikz}
\usepackage{pgfplots}
\usepackage{etoolbox}
\usepgfplotslibrary{patchplots}
\usepgfplotslibrary{colormaps}
\usepgfplotslibrary{groupplots}
\usepackage{algorithm2e}
\usepackage{soul}
\usepackage{enumerate}
\usetikzlibrary{positioning,arrows,mindmap,backgrounds, shapes, decorations.fractals, calc, intersections,decorations.pathmorphing,patterns}

\usepackage[colorinlistoftodos, obeyFinal]{todonotes}

\newtheorem{remark}{Remark}
\newcommand{\T}[1]{{\bm{#1}}}

\newcommand{\ie}{{i.e.,~}}
\newcommand{\eg}{{e.g.,~}}

\def\rmd{{\mathrm{d}}}
\def\be{\begin{equation}}
\def\ee{\end{equation}}
\def\ba{\begin{array}}
\def\ea{\end{array}}

\definecolor{dgreen}{RGB}{0,139,0}

	\title{A cost-effective isogeometric approach for composite plates based on a stress recovery procedure}
	\author[pavia1]{John-Eric Dufour\corref{cor1}}
	\author[lausanne]{Pablo Antolin}
	\author[pavia2,imati]{Giancarlo Sangalli}
	\author[pavia1,imati]{Ferdinando Auricchio}
	\author[pavia1,imati,TUM]{Alessandro Reali}

	\cortext[cor1]{Corresponding author}
	\address[pavia1]{Department of Civil Engineering and Architecture\\
		University of Pavia\\
		Via Ferrata, 3, 27100 Pavia, Italy\\
		e-mail: johnericpierre.dufour@unipv.it}
	
	\address[lausanne]{Institute of Mathematics\\
		\'Ecole Polytechnique F\'ed\'erale de Lausanne\\
		CH-1015 Lausanne, Switzerland}
	\date{\empty}

\address[pavia2]{Department of Mathematics\\
	University of Pavia\\
	via Ferrata 5, 27100 Pavia, Italy}
\address[imati]{Instituto di Matematica Applicata e Tecnologie Informatiche ``E.~Magenes'' (CNR)\\
Italy}
\address[TUM]{Technische Universit{\"a}t M{\"u}nchen\\
	Institute for Advanced Study\\
	Lichtenbergstra{\ss}e 2a, 85748 Garching, Germany}

\pgfplotsset{stress plot style/.style={%
  width=.49\textwidth,
  y label style={at={(axis description cs:0.15,.5)},anchor=south},
  ylabel={Thickness [mm]}}}

\newif\ifrecompiletikz
\recompiletikzfalse

\ifrecompiletikz
 \usetikzlibrary{external}
 \tikzexternaldisable
\fi

\begin{document}
		
	\begin{abstract}
    This paper introduces a cost-effective strategy to simulate the behavior of laminated plates by means of isogeometric 3D solid elements. Exploiting the high continuity of spline functions and their properties, a proper out-of-plane stress state is recovered from a coarse displacement solution using a post-processing step based on the enforcement of equilibrium in strong form. Appealing results are obtained and the method is shown to be particularly effective on slender composite stacks with a large number of layers. 		
	\end{abstract}

  \setcounter{tocdepth}{1}
	\maketitle
	\section{Introduction}
	Composite materials are used in a wide variety of fields such as aerospace or automotive. The study of composite laminates has become more and more important along the past years especially because of their light weight and very resistant mechanical properties. A composite laminate is usually made of several layers of highly resistant fibers embedded in soft matrix. Laminate structures tend to be prone to damage at the interfaces between layers, this mode of failure being referred to as delamination. The prediction and evaluation of damage in composite laminates demands an accurate evaluation of the three-dimensional stress state through the thickness, although most of the studies available in the literature consider the laminate as a two-dimensional object. Classical two-dimensional theories such as shell approaches are not accurate enough to reliably predict interlaminar damage and delamination. Accordingly, a number of layerwise theories have been developed to compute more accurately the mechanical state inside the laminate. Such methods often rely on heavy computations and hybrid approaches in order to be able to capture the complex behaviour of the interlaminar interfaces. The numerical counterpart of such layerwise or hybrid theories was approached mostly using standard finite element discretization (see, \eg~\cite{Reddy2004,Carrera2002} and references therein).
	
	
	 Over the last decade, many novel methods have been proposed to enhance the standard finite element framework. Among them, Isogeometric Analysis (IGA) is a concept proposed by Hughes et al.~\cite{Hughes2005} where the shape functions used in computer aided design are also approximating physical fields and state variables. It thus relies on spline functions, like, \eg NURBS (Non Uniform Rational B-Splines), to approximate both the geometry and the solution. In addition to improve the pipeline between design and analysis, spline shape functions possess properties which can be used to drastically improve the performance of numerical analysis compared to standard finite element method. For example, spline shape functions can provide easily high order approximation and highly simplify the refinement process. Moreover, their smoothness guarantees higher accuracy and opens the door to the discretization of high-order PDEs in primal form. IGA has been successfully used to tackle a large variety of problems, including solid and structures ( see, \eg~\cite{Borden2012,Cottrell2006,Elguedj2008,Morganti2015,Kiendl2009}), fluids (see, \eg~\cite{Akkerman2008,Bazilevs2007}), fluid-structure interaction (see, \eg~\cite{Bazilevs2011,Hsu2015}), and other fields (see \eg~\cite{Gomez2008,Lorenzo2017}).
	 
	IGA methods have already been used to solve composite laminate problems. A wide variety of laminated models have been implemented and studied within the IGA framework, especially relying on high-order theories~\cite{Thai2015,Kapoor2013,Remmers2015}. These techniques mostly utilize enhanced shell and plate theories. In addition to these 2D approaches, some methods can be used to compute the full 3D stress state using 3D isogeometric analysis such as in~\cite{Remmers2015,Guo2014,Guo2015}. In such approaches, each ply of the laminate is delimited by a C$^{0}$-continuous interface. Such an approach (referred in the following as ``layerwise'' ) involves a large number of degrees of freedom when a lot of layers are composing the laminate. Thus it may be not completely satisfactory, despite its accuracy, due to its high cost.
	
	This paper presents an approach consisting in using 3D computations with a reduced number (namely, one) of elements through the thickness and a layerwise integration rule, which is then post-processed in order to obtain an accurate 3D stress state. The post-processing relies on the integration of the equilibrium equations (such as in the recovery techniques proposed in~\cite{Kapoor2013,Pryor1971,Engblom1985,Daghia2008,Ubertini2004,Daghia2013,deMiranda2002}) to compute the stresses through the thickness from the in-plane ones. This allows to drastically reduce the computational time compared to the 3D layerwise approach as the number of degrees of freedom is significantly decreased. The post-processing operation is in fact very fast and its cost does not increase significantly with the number of degrees of freedom. Solutions obtained using the proposed technique are actually close to those provided by full ``layerwise'' 3D isogeometric analysis, at a fraction of the cost, in particular when many layers are present.
	
	The effectiveness of the proposed approach relies on two ingredients, both granted by the peculiar properties of IGA, namely:
	\begin{enumerate}[(i)]
		\item The capability of obtaining accurate in-plane results with a coarse mesh with only one element through the thickness;
		\item The higher continuity allowing an accurate computation of stresses and of their derivatives from displacements.
	\end{enumerate}
	The structure of the paper is as follows.
	First, the considered isogeometric strategies are briefly introduced. Then, the 3D geometry and the mechanical problem used as a test case is detailed along with its analytical solution. Results using the two considered standard isogeometric approaches are then presented. Finally, the post-processing approach is proposed along with numerical results and a comparison with those provided by full ``layerwise'' 3D isogeometric analysis. Several cases using different numbers of quadrature points per material layer, numbers of layers and thicknesses are considered to show the effectiveness of the method.
		
	\section{Standard IGA strategies for 3D analysis of laminates}
\begin{figure}
		\centering
    \subfigure[Layerwise approach\label{subfig-1:shapefunction}]{\ifrecompiletikz\tikzsetnextfilename{fig_01_a}\tikzexternalenable\input{Images/fig_01_a}\tikzexternaldisable\else\includegraphics{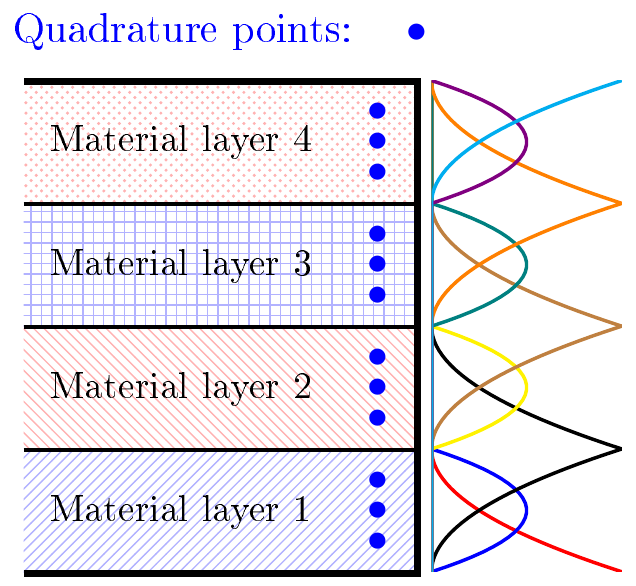}\fi}\hfill
    \subfigure[Single-element approach\label{subfig-2:shapefunction}]{\ifrecompiletikz\tikzsetnextfilename{fig_01_b}\tikzexternalenable\input{Images/fig_01_b}\tikzexternaldisable\else\includegraphics{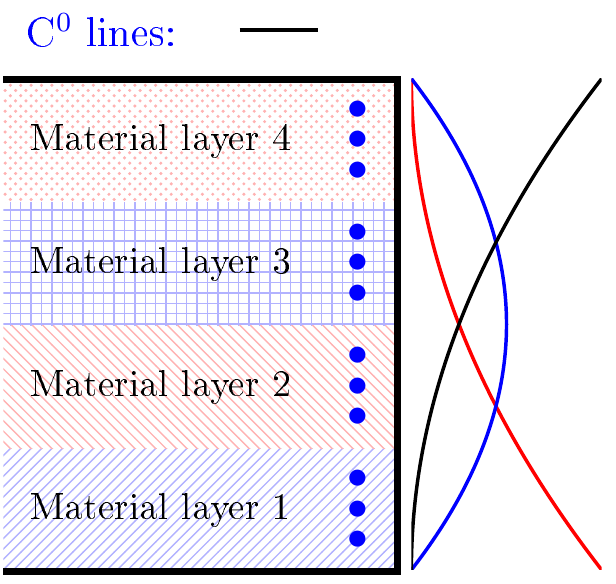}\fi}
		\caption{Isogeometric shape functions used in the two considered approaches}
		\label{fig:shapefunction}
	\end{figure}

	In this section we present a preliminary discussion on standard IGA strategies for 3D analysis of laminates. For the sake of conciseness we avoid reporting an introduction on the basic concepts of IGA for 3D elasticity, including basics of B-Splines, NURBS, etc. Readers are instead refered to \cite{Cottrell09} and references herein. Here, we just recall that we will consider a standard IGA consisting of an isoparametric Galerkin formulation where splines are used for approximating both geometry and displacements.
	As opposed to the framework typically used to compute laminate shell and plate solutions~\cite{Reddy2004,Kapoor2013,Barbero1991}, we choose here to perform a full ``layerwise'' 3D computation.
	Such an approach has already been used in combination with IGA in order to solve multilayered plate problems~\cite{Remmers2015,Guo2014,Guo2015} and along those lines we start considering  two different sets of shape functions, as follows:\newline
	\textit{Layerwise approach:} The first considered approach is similar to high order finite element methods. Each layer is in fact modeled by one patch through the thickness and $C^0$ continuity is kept between elements~(see Figure~\ref{subfig-1:shapefunction}). Such an approach is completely layerwise and, as a consequence, the number of degrees of freedom is proportional to the number of layers. A standard integration rule is adopted, namely $p$ + 1 points in each direction ($p$ being the degree of the shape functions). This approach is equivalent to the one proposed in~\cite{Remmers2015,Guo2015}.\\
	\textit{Single-element approach:} The second approach actually uses a single element
	 through the thickness (see Figure \ref{subfig-2:shapefunction}), strongly reducing the number of degrees of freedom with respect to the previously mentioned method. To account for the presence of the layers, a special integration rule is adopted, consisting of a $q$-point Gauss rule ($q\geq1$) over each layer. This can be considered as a special homogenized approach with a layerwise integration. Such a method would \textit{a priori} not give sufficiently good results (in terms of through-the-thickness stress description) but could be easily coupled with a post-processor to improve the solution, and this will be the object of Section~\ref{sec:recovery}.\\
	It should be noted that the in-plane continuity of the shape functions in both approaches is the same.
	
	\section{Test case: The Pagano layered plate}
	
\begin{figure}
		\centering
		\includegraphics[width=.8\textwidth]{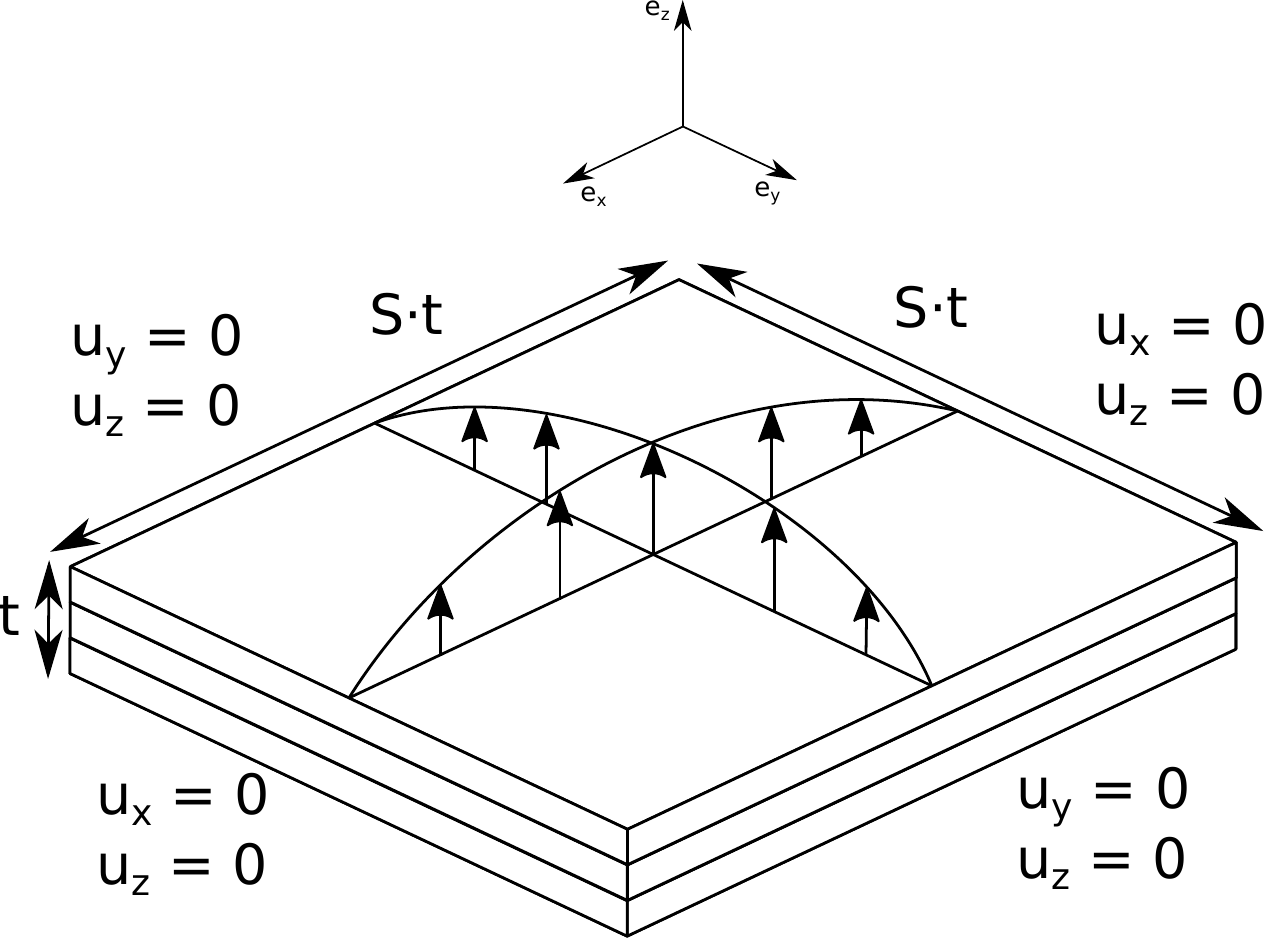}
		\caption{3D problem used as a test case, proposed by Pagano~\cite{Pagano1970}}
		\label{fig:testproblem}
	\end{figure}

  The test case used in this study is the classical one proposed by Pagano~\cite{Pagano1970}. It consists of a simply supported multilayered 3D plate with a sinusoidal loading on top and a loading-free bottom face (see Figure \ref{fig:testproblem}). This problem can be easily parameterized which allows to analyze many cases in terms of layer number and distribution.
  In the following study, a few examples are considered using different numbers of layers (\ie 3, 4, 11, and 34). In all these cases, the loading conditions are the same (namely, a two dimensional sinus with a period equal to twice the length of the plate), while the thickness of every single layer is set to $1\,\text{mm}$, and the length of the plate is chosen to be $S$ times larger than the total thickness $t$ of the laminate. Figure~\ref{fig:testproblem} shows the summary of the test problem in the case of three layers.
	
	The laminate is composed by orthotropic layers placed orthogonally on top of each other (thus creating a 90/0/90/... laminate). 
	For each layer, we have:
    \begin{equation}\label{eq:constitutive}
    \T{\sigma} = \mathbb{C}\T{\varepsilon}
	\end{equation}
  where the elasticity tensor $\mathbb{C}$ can be expressed, using Voigt notation, as:
	\begin{equation}
  \mathbb{C}=\left[\begin{matrix}
	\dfrac{1}{E_1} & -\dfrac{\nu_{12}}{E_2} & -\dfrac{\nu_{13}}{E_1} & 	0 & 	0 & 0 \\ 
	-\dfrac{\nu_{21}}{E_2} & \dfrac{1}{E_2} & -\dfrac{\nu_{23}}{E_2} & 	0 & 	0 & 0 \\ 
	-\dfrac{\nu_{31}}{E_3} & -\dfrac{\nu_{32}}{E_3} & \dfrac{1}{E_3} & 	0 & 	0 & 0 \\ 
	0 &		 0 &	0 & \dfrac{1}{G_{23}} & 	0 & 0 \\ 
	0 &		 0 & 	0 & 	0 & \dfrac{1}{G_{31}} & 0  \\ 
	0 &		 0 & 	0 & 	0 & 	0  & \dfrac{1}{G_{12}}
	\end{matrix} \right]^{-1}\,,
	\end{equation}
and the material parameters for these layers are the following:
	\begin{equation}
	\begin{aligned}
	    E_2 &=  E_3 =  \frac{E_1}{25}  \\
      G_{12} &= G_{13} = \frac{G_{23}}{2.5} \\
		\nu_{12} &= \nu_{13} = \nu_{23} = 0.25
	\end{aligned}
	\end{equation}

	The pressure field used as the top boundary condition is
	\begin{equation}
		p(x,y) = \sigma_0 \sin\left(\frac{\pi x}{S t}\right) \sin\left(\frac{\pi y}{S t}\right)
	\end{equation}
	In our test, we always select $E_1=25$~GPa, $G_{23}=0.5$~GPa and $\sigma_0=1$.
	\subsection{Analytical results}
	Each stress component presents a different distribution along thickness.
	In particular, in-plane normal components of the stress tensor (\ie $\sigma_{11}$ and $\sigma_{22}$) are discontinuous along the thickness although the displacement solution is continuous in all cases. 
  This is obviously due to the differences in term of material parameters between the layers. On the other hand, equilibrium implies that through-the-thickness stress components have to be C$^{0}$-continuous (i.e., $\sigma_{13}$, $\sigma_{23}$ and $\sigma_{33}$)
  .
  Along the article, stress components are normalized, according to \cite{Pagano1970}, using the following formulas:
	\begin{equation}
		\begin{aligned}
		\bar{\sigma}_{ij} &= \frac{\sigma_{ij}}{\sigma_0 S^2}  &i=1,2 \quad j=1,2 \\
		\bar{\sigma}_{i3} &= \frac{\sigma_{i3}}{\sigma_0 S}  &i=1,2 \\
		\bar{\sigma}_{33} &= \frac{\sigma_{33}}{\sigma_0}
		\end{aligned}
	\end{equation}
	
	Figure~\ref{fig:example_sig} reports the analytical solution (computed using the equations given in \cite{Pagano1970}) for the  $\sigma_{13}$, $\sigma_{11}$ and $\sigma_{12}$  components of the stress tensor in the case of 3 and 11 layers (blue solid lines). We can clearly identify the different behavior of the layers, with a ``soft'' and a ``hard'' direction.
	
	\subsection{First results}
	
\begin{figure}
		\centering
    \subfigure[$\sigma_{13}$\label{subfig-1:test3}]{\ifrecompiletikz\tikzsetnextfilename{fig_03_a}\tikzexternalenable\input{Images/fig_03_a}\tikzexternaldisable\else\includegraphics{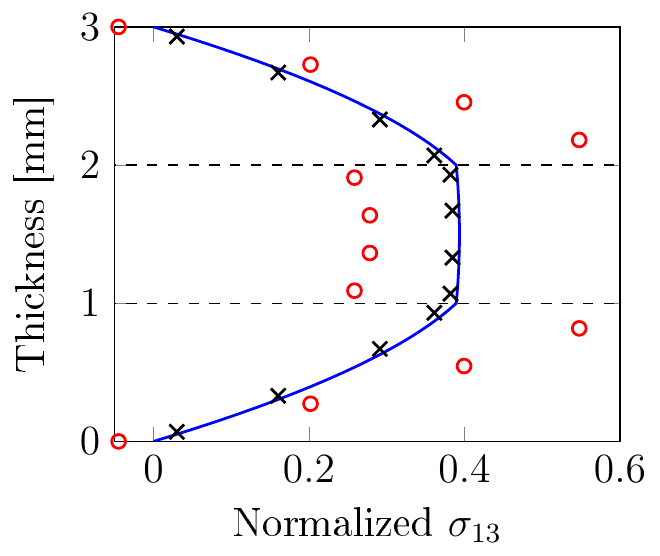}\fi}\hfill
    \subfigure[$\sigma_{13}$\label{subfig-1:test11}]{\ifrecompiletikz\tikzsetnextfilename{fig_04_a}\tikzexternalenable\input{Images/fig_04_a}\tikzexternaldisable\else\includegraphics{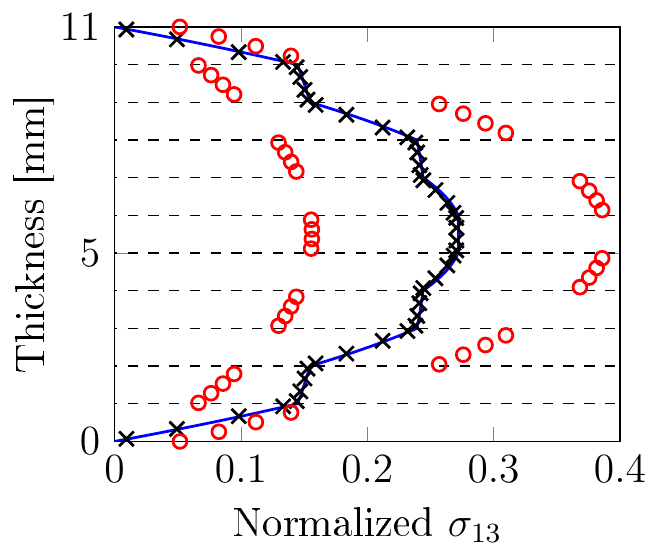}\fi}\\
    \subfigure[$\sigma_{11}$\label{subfig-2:test3}]{\ifrecompiletikz\tikzsetnextfilename{fig_03_b}\tikzexternalenable\input{Images/fig_03_b}\tikzexternaldisable\else\includegraphics{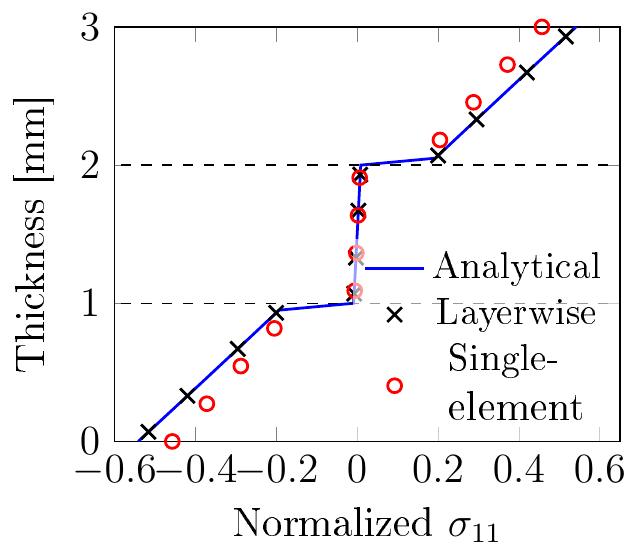}\fi}\hfill
    \subfigure[$\sigma_{11}$\label{subfig-2:test11}]{\ifrecompiletikz\tikzsetnextfilename{fig_04_b}\tikzexternalenable\input{Images/fig_04_b}\tikzexternaldisable\else\includegraphics{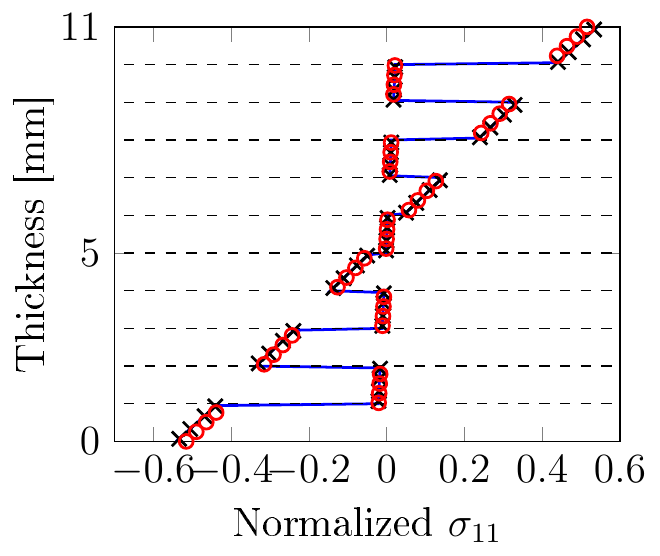}\fi}\\
    \subfigure[$\sigma_{12}$\label{subfig-3:test3}]{\ifrecompiletikz\tikzsetnextfilename{fig_03_c}\tikzexternalenable\input{Images/fig_03_c}\tikzexternaldisable\else\includegraphics{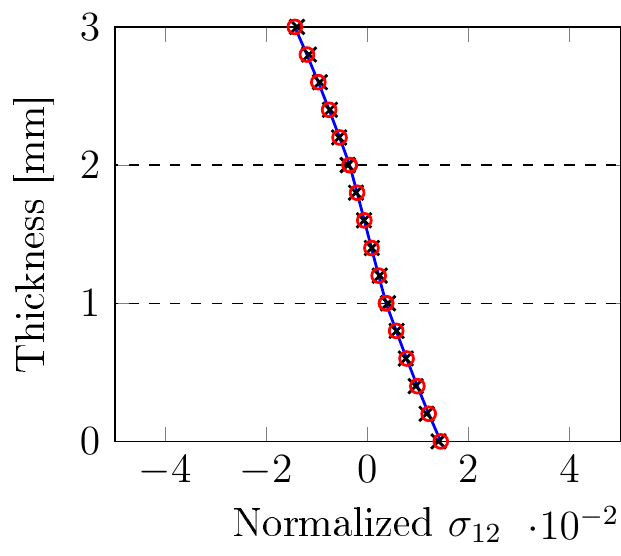}\fi}\hfill
    \subfigure[$\sigma_{12}$\label{subfig-3:test11}]{\ifrecompiletikz\tikzsetnextfilename{fig_04_c}\tikzexternalenable\input{Images/fig_04_c}\tikzexternaldisable\else\includegraphics{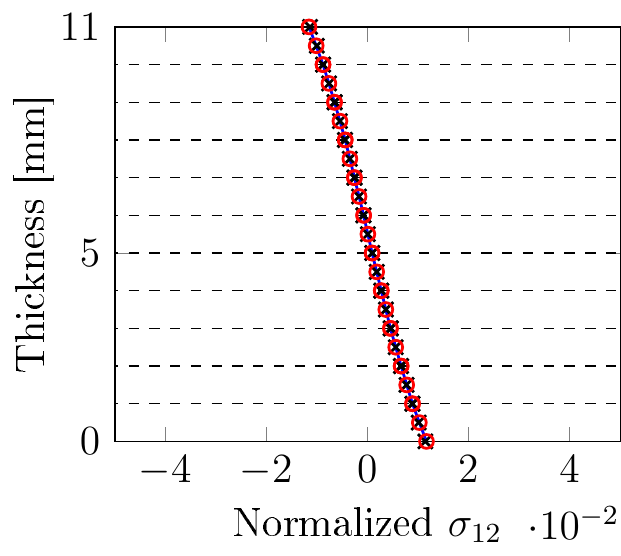}\fi}
		\caption{Computed solutions for the 3D Pagano plate problem with 3 and 11 layers at the position $X=0.25L$ and $Y=0.25L$. The blue solid line represents the analytical solution, the black cross represent the solution using the layerwise approach, and the red circles represent the solution using the single-element approach (without post-processing)}
		\label{fig:example_sig}
	\end{figure}

	In this section, we comment the results obtained using the layerwise and the single-element approaches, which are compared with the analytical solutions for $\sigma_{13}$, $\sigma_{11}$ and $\sigma_{12}$ in Figure~\ref{fig:example_sig} for the cases of 3 and 11 layers at the position $X=0.25L$ and $Y=0.25L$. Results are given for an in-plane mesh composed of $9\times 9$ quartic elements, with cubic shape functions through the thickness.
	All the results shown in this work have been obtained using an in-house code based on the igatools library (see~\cite{Pauletti2015} for further details).
%
	
	The layerwise approach captures accurately the solution for both displacements and stresses, no matter if they are discontinuous or not (see Figure~\ref{fig:example_sig}). This is due to the layerwise displacement $C^0$-continuity which allows a proper description of the stress state even between two layers.
	
	On the contrary, using a poor description (\ie a single high order element through the thickness) does not allow a correct description of the stress state through the thickness (see Figure~\ref{fig:example_sig}). The high order continuity of the shape functions leads in fact to the continuity of strains and the jump in the material parameters from one layer to another leads to the discontinuity of the stress components (even when they should be continuous through the thickness). Therefore, the differences between the analytical solution and the computed results are very significant.

	\begin{figure}
		\centering
    \ifrecompiletikz\tikzsetnextfilename{fig_05}\tikzexternalenable\input{Images/fig_05}\tikzexternaldisable\else\includegraphics{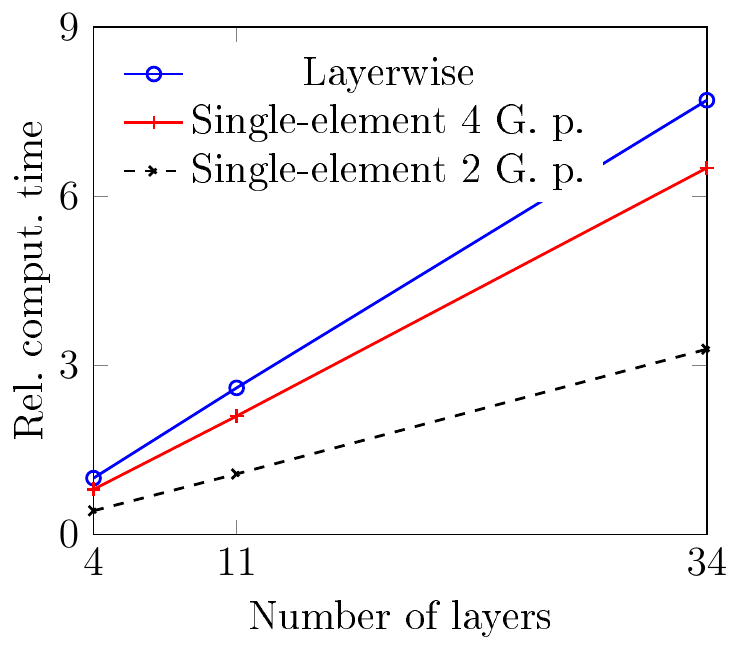}\fi
		\caption{Relative computational time with respect to the number of layers. The single-element approach (red and black lines, using $4$ and $2$ integration points  per layer respectively) is less time consuming than the simulation with one element per layer (blue line). In this case, there are 9 quartic elements along each in-plane direction and cubic shape functions through the thickness, corresponding to 1,183, 5,746 and 17,407 degrees of freedom for the layerwise approach, and 676 for the single-element approach
    }
    \label{fig:comptime}
	\end{figure}

	Figure~\ref{fig:comptime} gathers the relative computation time for these two approaches with different numbers of layers. As expected, the computational time of the layerwise approach is higher because the number of degrees of freedom is higher (1,183, 5,746 and 17,407 versus 676) as well as the number of integration points required per layer. Note that the computational times of the layerwise approach are, in this case, about 2.5 times higher than the single-element approach with 2 Gauss points per layer.
 
	The layerwise approach seems to be suitable for small stacks (\ie with a small amount of layers), while it is not suitable for high number of layers as it becomes extremely time and memory consuming.
	
	On the other hand, the single-element approach seems to be more scalable, as only one element through-the-thickness is used, but only the global solution in terms of displacement and in-plane stresses is accurately captured by the simulation. The idea of the paper is to obtain a better approximation of the out-of-plane stresses by post-processing the accurately computed displacements and in-plane stresses, without significantly increasing the overall simulation cost.
	
	\section{Reconstruction from Equilibrium}
	\label{sec:recovery}
  Although in-plane stress components are almost correctly captured by the single-element approach (as it can be seen in Figures \ref{subfig-2:test3} and \ref{subfig-2:test11} for the component $\sigma_{11}$ and \ref{subfig-3:test3} and \ref{subfig-3:test11} for the component $\sigma_{12}$), the out-of-plane components are not. As interlaminar delamination and other fracture processes rely mostly on such out-of-plane components, a proper through-the-thickness stress description is required. In order to compute a more accurate stress state, we choose to use the following post-processing approach based on the equilibrium equations (as it has been proposed in \cite{Pryor1971,Engblom1985,Ubertini2004,deMiranda2002} in a finite element framework), relying on the higher regularity granted by IGA shape functions.

	In an equilibrium state, the stresses inside the material should satisfy the equilibrium equation
	\begin{equation}
	\T{\nabla} \cdot \T{\sigma} = \T{b}
	\end{equation}
	where $\T{\nabla}\cdot$ is the divergence operator.
	In a 3D case, this reads in components as
	\begin{equation}
		\begin{aligned}
			\sigma_{11,1} + \sigma_{12,2} + \sigma_{13,3} &= b_1\\
			\sigma_{12,1} + \sigma_{22,2} + \sigma_{23,3} &= b_2\\
			\sigma_{13,1} + \sigma_{23,2} + \sigma_{33,3} &= b_3
		\end{aligned}
	\end{equation}
	with \[\sigma_{ij,k}=\frac{\partial \sigma_{ij}}{\partial x_k}\,.\]
	By integrating along the thickness, we recover the out-of-plane stresses as
	%
  \begin{subequations}
  \begin{align}
  \label{eq:s13}
	\sigma_{13}(z) &= -\int^{z}_{z_0}(\sigma_{11,1} + \sigma_{12,2}-b_1)\rmd z + \sigma_{13}(z_0)\,, \\
  \label{eq:s23}
	\sigma_{23}(z) &= -\int^{z}_{z_0}(\sigma_{12,1} + \sigma_{22,2}-b_2)\rmd z + \sigma_{23}(z_0)\,, \\
  \label{eq:s33}
	\sigma_{33}(z) &= -\int^{z}_{z_0}(\sigma_{13,1} + \sigma_{23,2}-b_3)\rmd z + \sigma_{33}(z_0)\,,
  \end{align}
  \end{subequations}
    It is important to note that, by introducing equations \eqref{eq:s13} and \eqref{eq:s23} into \eqref{eq:s33}, the component $\sigma_{33}$ can then be computed as
	\begin{equation}
		\sigma_{33}(z) = \int^{z}_{z_0}(\sigma_{11,11} + \sigma_{22,22} + 2\sigma_{12,12} + b_3 - b_{1,1}-b_{2,2})\rmd z + \sigma_{33}(z_0)\,.
	\end{equation}
  The integral constants should be chosen to fulfill the boundary conditions at the top or bottom surfaces.
  \begin{remark}
  	 It should be noted that the integration could be performed from both surfaces and then averaged in order to divide the resulting error by two. Although it is easy to compute the boundary conditions in this case, we chose to avoid this as it relies on the perfect knowledge of the stress boundary conditions on both surfaces (including the loaded one) which are not always available.
  \end{remark}
  Assuming that the elasticity tensor $\mathbb{C}$ of equation \eqref{eq:constitutive} is constant, the derivatives of the in-plane components of the stress are computed from displacements as
	\begin{equation}
	\begin{aligned}
    \sigma_{ij,k} &= \mathbb{C}_{ijmn} \varepsilon_{mn,k}\,, \\
		\sigma_{ij,kl} &= \mathbb{C}_{ijmn} \varepsilon_{mn,kl}\,,
	\end{aligned}
	\end{equation}
	where
	\begin{equation}
	\begin{aligned}
	\varepsilon_{mn,k} &= \frac{1}{2} (u_{m,nk}+u_{n,mk})\,,\\
	\varepsilon_{mn,kl} &= \frac{1}{2} (u_{m,nkl}+u_{n,mkl})\,.
	\end{aligned}
	\label{eq:prereq}
	\end{equation}
	
	Equation~\eqref{eq:prereq} clearly demonstrates the necessity of a highly regular displacement solution in order to recover a proper stress state. Such condition can be easily achieved using IGA with B-spline shape functions of order $p>2$ while in a standard finite element framework, continuity of the solution must be enhanced (see~\cite{Ubertini2004,deMiranda2002}).
	One could thus retrieve a good approximation of the out-of-plane stress state once an accurate description of the in-plane stress state is available. 
	
	The displacement solution computed using an IGA method is a spline, which means that its post-processing and further handling of its derivatives can be efficiently performed. Splines can be indeed easily differentiated and many dedicated tools are available~\cite{Piegl1997}.

\subsection{Numerical results}
	\begin{figure}
		\centering
    \ifrecompiletikz\tikzsetnextfilename{fig_06}\tikzexternalenable\input{Images/fig_06}\tikzexternaldisable\else\includegraphics{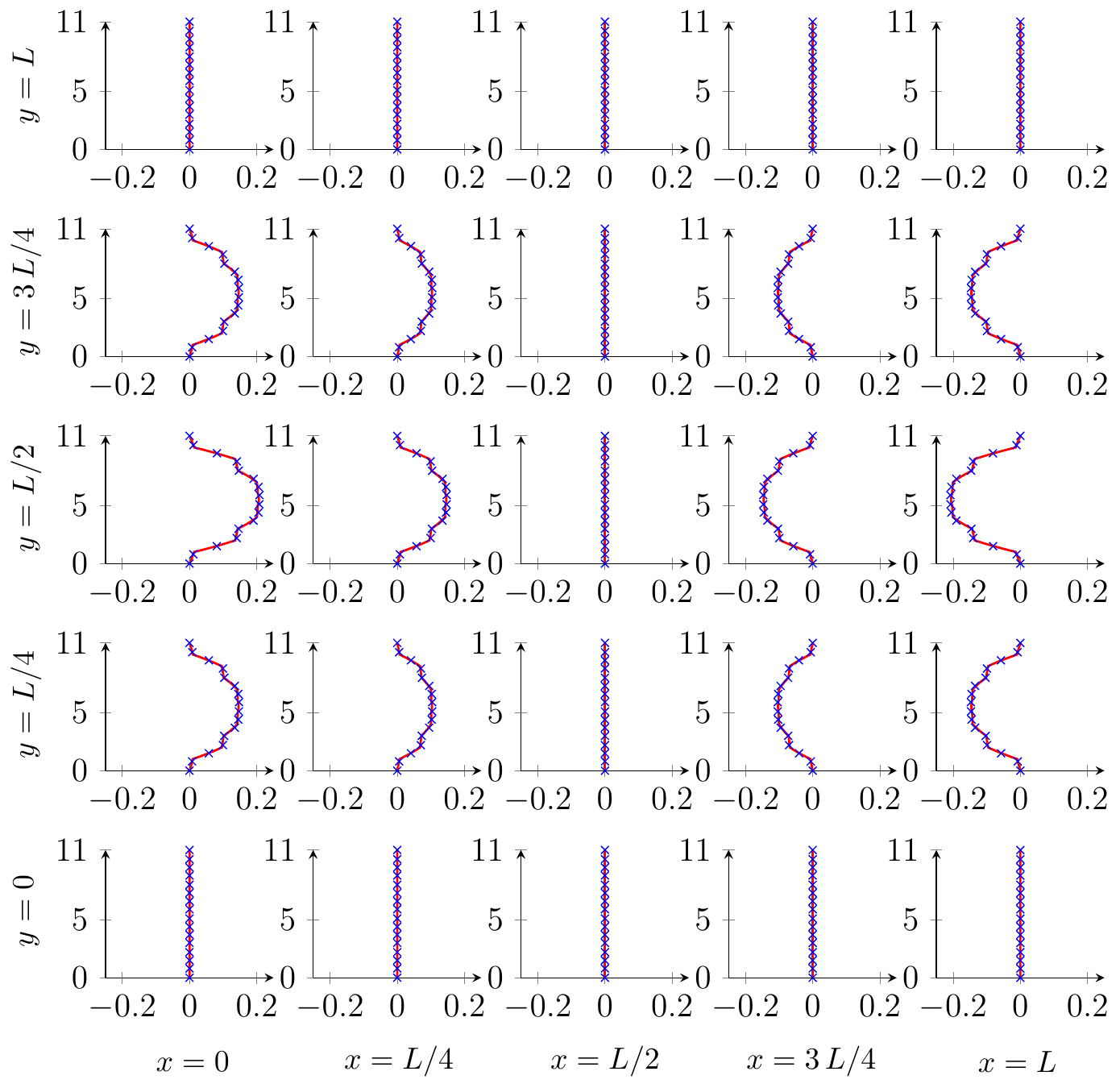}\fi
		\caption{Recovered (red solid line) $\sigma_{13}$ compared to the analytical one (blue crosses) for several in plane positions.
    $L$ is the total length of the plate, that in this case is  $L=110\,\text{mm}$ (being $L=S\,t$ with $t=11\,\text{mm}$ and $S=10$), while the number of layers is 11.}
		\label{fig:recov1}
	\end{figure}

	\begin{figure}
		\centering
    \ifrecompiletikz\tikzsetnextfilename{fig_07}\tikzexternalenable\input{Images/fig_07}\tikzexternaldisable\else\includegraphics{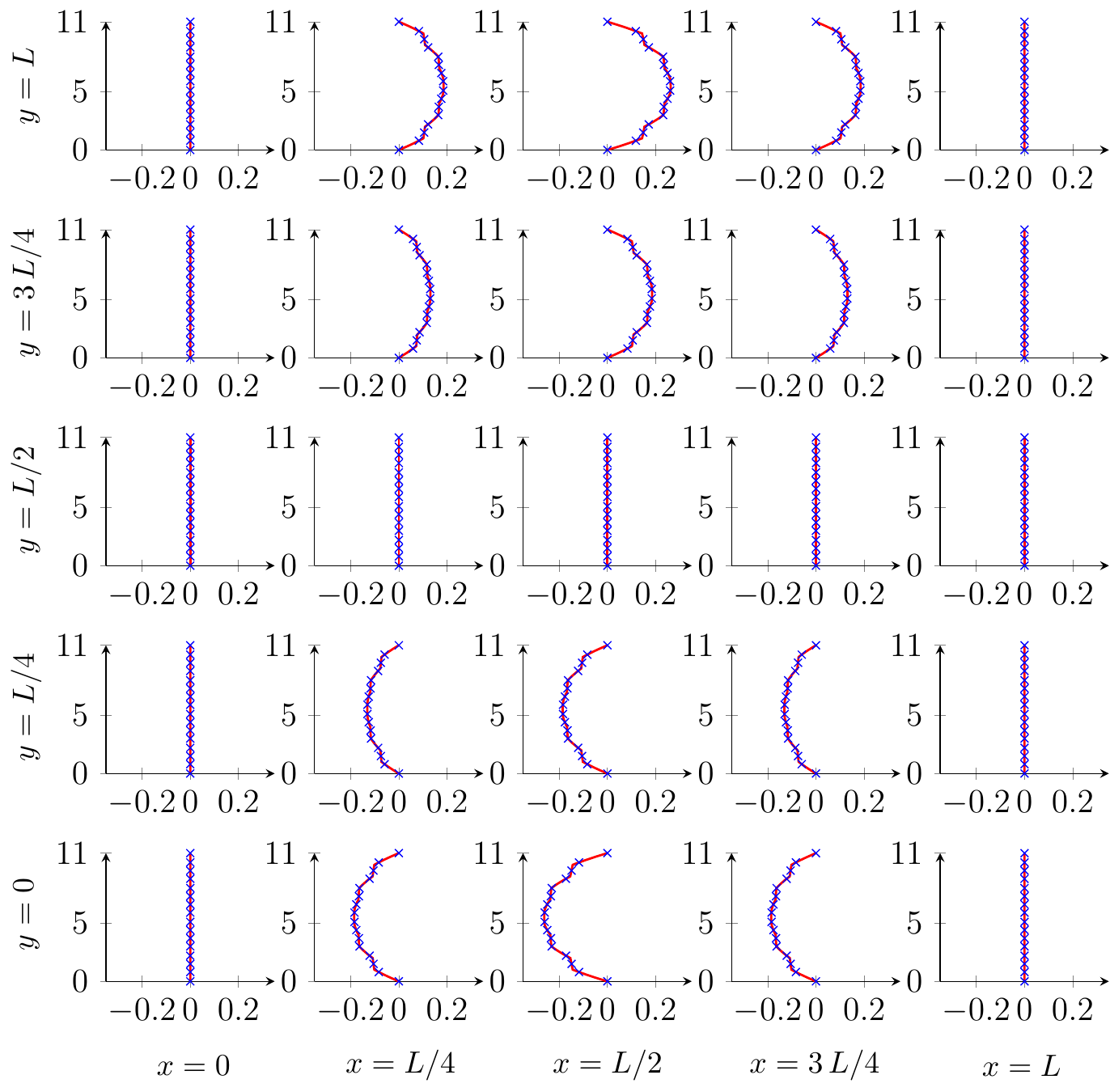}\fi
		\caption{Recovered (red solid line) $\sigma_{23}$ compared to the analytical one (blue crosses) for several in plane positions.
    $L$ is the total length of the plate, that in this case is  $L=110\,\text{mm}$ (being $L=S\,t$ with $t=11\,\text{mm}$ and $S=10$), while the number of layers is 11.}
		\label{fig:recov2}
	\end{figure}
\begin{figure}
	\centering
  \ifrecompiletikz\tikzsetnextfilename{fig_08}\tikzexternalenable\input{Images/fig_08}\tikzexternaldisable\else\includegraphics{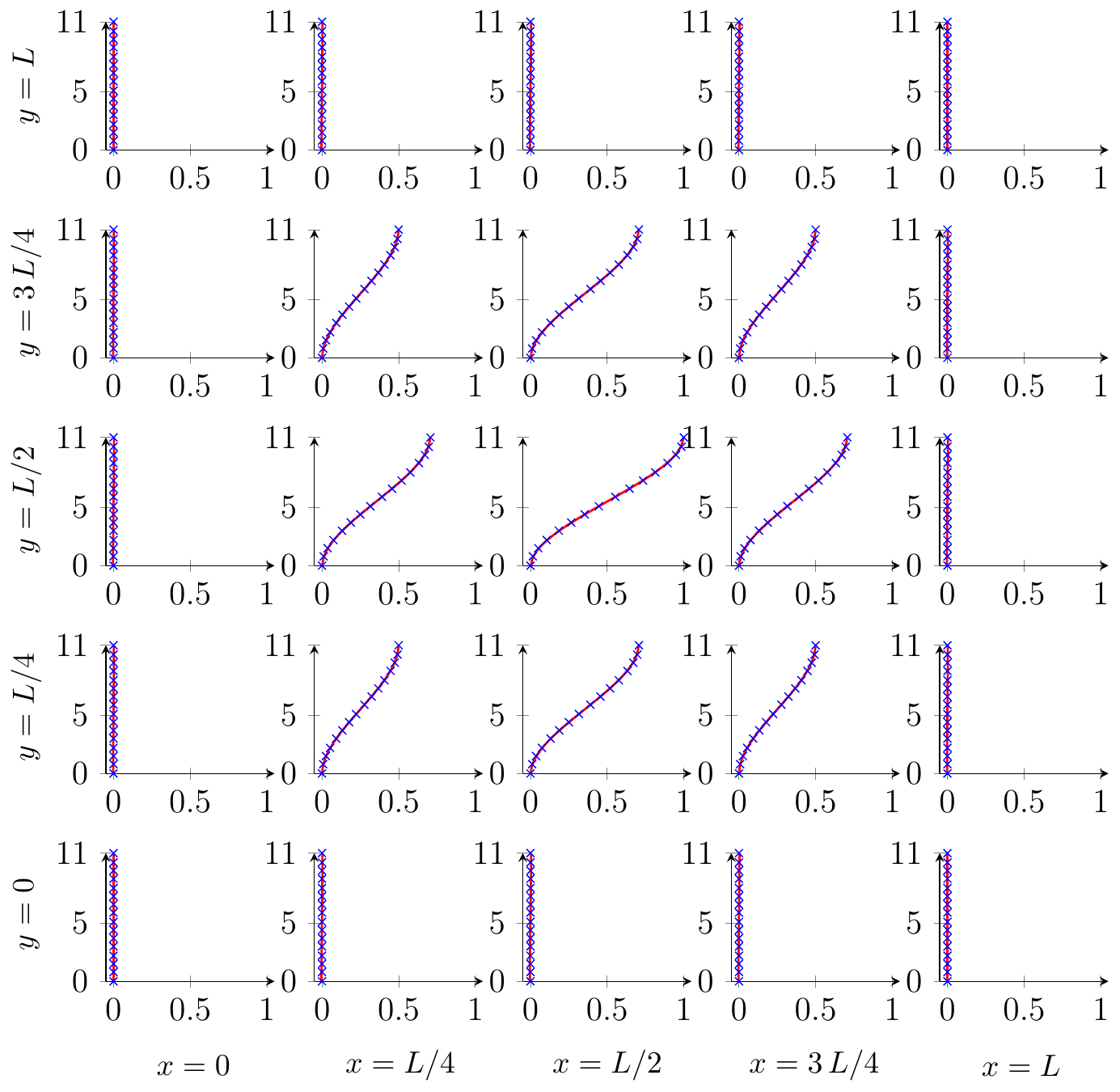}\fi
	\caption{Recovered (red solid line) $\sigma_{33}$ compared to the analytical one (blue crosses) for several in plane positions.
    $L$ is the total length of the plate, that in this case is  $L=110\,\text{mm}$ (being $L=S\,t$ with $t=11\,\text{mm}$ and $S=10$), while the number of layers is 11.}
	\label{fig:recov3}
\end{figure}

	Figures~\ref{fig:recov1},~\ref{fig:recov2} and~\ref{fig:recov3} show that applying the post-processing allows to greatly improve the quality of the results. Indeed, we observe that the post-processed stress state is very close to the analytical one. The correct stress state is recovered on the whole plate, and a great improvement is observed everywhere. The results were obtained using an in-plane mesh composed of $9\times 9$ quartic elements with cubic shape functions through the thickness.
	
\begin{figure}
		\centering
    \subfigure[4 layers\label{subfig-1:errors}]{\ifrecompiletikz\tikzsetnextfilename{fig_09_a}\tikzexternalenable\input{Images/fig_09_a}\tikzexternaldisable\else\includegraphics{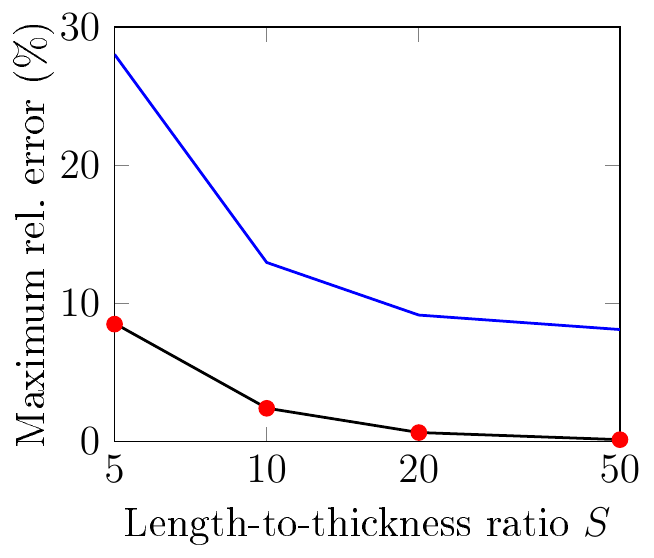}\fi}\hfill
    \subfigure[11 layers\label{subfig-2:errors}]{\ifrecompiletikz\tikzsetnextfilename{fig_09_b}\tikzexternalenable\input{Images/fig_09_b}\tikzexternaldisable\else\includegraphics{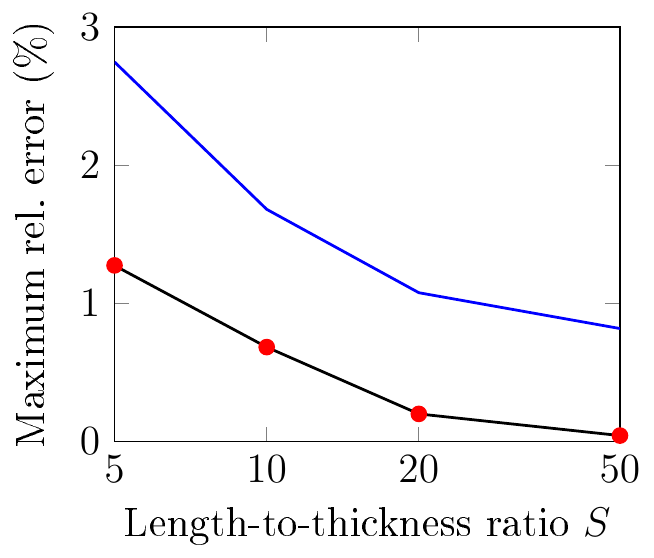}\fi}\\
    \subfigure[34 layers\label{subfig-3:errors}]{\ifrecompiletikz\tikzsetnextfilename{fig_09_c}\tikzexternalenable\input{Images/fig_09_c}\tikzexternaldisable\else\includegraphics{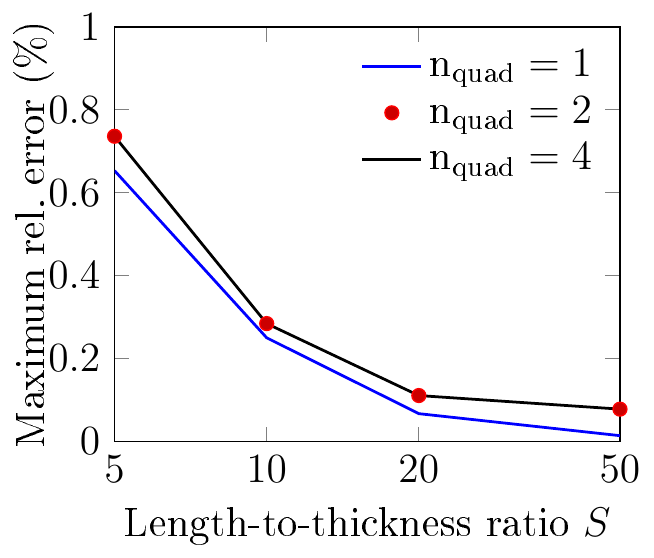}\fi}
		\caption{Difference between post-processed and analytical stress state. Different numbers of integration points per layer along the thickness ($n_{\text{quad}}$) and length-to-thickness ratios $S$ are used.}
		\label{fig:results}
	\end{figure}

  In order to validate such an approach in a wider variety of cases, computations with several sets of number of layers for the laminate or a different ratio between the thickness of the plate and its length can be performed. Figure~\ref{fig:results} gathers the results for different cases (varying the number of layers) with respect to the length-thickness ratio $S$ used in the computations. In each layer, a standard Gaussian integration with $p + 1$ points (\ie 5, in our case) along each in-plane direction has been used. The error is computed as:
	\begin{equation}
		\text{error}=\frac{\max(\sigma_\text{analytic}-\sigma_\text{recov})}{\max(\sigma_\text{analytic})}\,.
	\end{equation}
	
  Several conclusions can be drawn from Figure~\ref{fig:results}. First, the post-processing method provides better results when the thickness ratio is larger. The method should therefore be used on slender plates rather than on thick stacks. Concerning the integration scheme, $p + 1$ points per layer provide basically the same accuracy as using $p - 1$ points per layer (\ie 2 points, given the cubic through-the-thickness approximation). $p-2$ integration points per layer (corresponding to just 1 point in our case) lead in general to a loss of accuracy except, in our example, in the case of 34 layers (Figure~\ref{subfig-3:errors}) where some super-convergence effect is present. Based on our tests, this ``super-convergence'' seems to be a special case happening only when a large unsymmetrical stack is involved.
  
  \begin{figure}
  	\centering
  	\ifrecompiletikz\tikzsetnextfilename{fig_13}\tikzexternalenable\input{Images/fig_13}\tikzexternaldisable\else\includegraphics{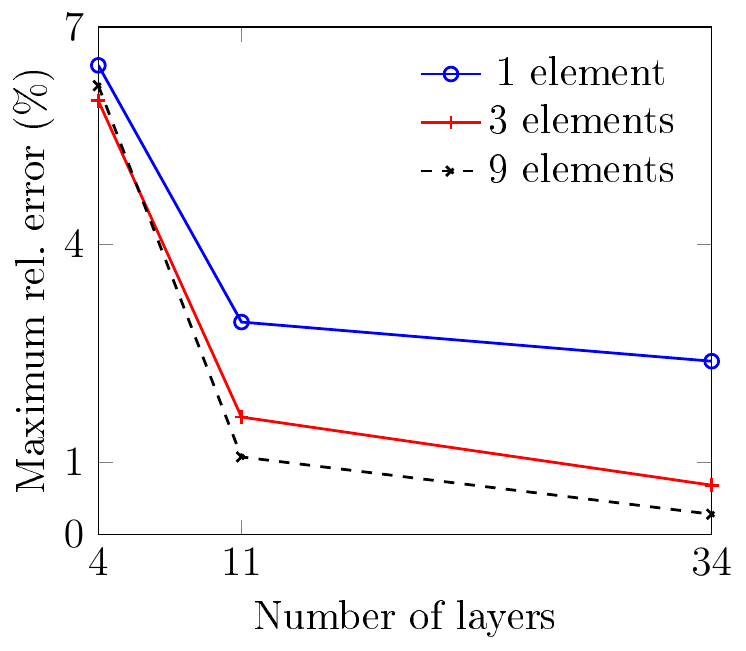}\fi
  	\caption{Difference between post-processed and analytical stress state. Different number of elements are used for the in-plane mesh, with $S = 10$. The degree of the shape functions remains 4 in-plane and 3 through the thickness\label{fig:error_elem}}
  \end{figure}

	Finally, the more layers in the stack, the better the results, since a stack with a large number of thin layers is indeed closer to a plate with average properties, whose solution would be perfectly captured by a standard isogeometric simulation with one high-order element through the thickness.

	Therefore, the post-processing approach seems to be particularly suited to handle large and thin plates with a significant number of layers (whereas the first approach with one element per layer is far too costly to handle these kinds of problems, due to the huge number of elements and degrees of freedom). In such situations, this method provides results very close to the correct solution, at a fraction of the cost, representing a far better choice than a full 3D computation in terms of accuracy-to-cost ratio.
	
	Although the results in this paper are obtained using a 9x9 elements in-plane mesh, Figure~\ref{fig:error_elem} shows that even using a single element mesh, the error remains acceptable in this case.

	\section{Conclusion}
	
	In this paper, we have presented a very simple approach for an accurate simulation of laminates combining a standard 3D coarse isogeometric analysis with a layerwise integration rule and a post-processing based on equilibrium equations. Using this approach, results close to the analytical solution are obtained, while using only few degrees of freedom through the thickness in contrast with a standard 3D layerwise approach.
	
	Multiple numbers of layers (both even and odd) have been considered in our numerical tests and different thicknesses and numbers of integration points per layer have been studied. Whatever the number of layers, the method gives better results the thinner the layers are. Based on our results, a through-the-thickness integration rule involving $p-1$ points per layer can be considered as the default choice, since it grants a good accuracy even when the plate becomes thicker.
	
	This method is easy to implement. Indeed, the post-processing is only based on the integration of equilibrium equations, and all the required components can be easily computed from the displacement solution. Furthermore, a standard 3D isogeometric computation is used, with a special set of integration points, but without any modification in the model. As the IGA code provides spline fields, its derivation and handling are straightforward and the whole process (coarse simulation plus post-processing) is far less time consuming than a full layerwise 3D approach.
	
	Further research topics currently under investigation involve the extension of this approach to more complex problems involving curved geometries and/or large deformations.
	Moreover, more efficient through-the-thickness integration strategies or the combination with bivariate shells will be also considered in future studies.
	\section*{Acknowledgment}
	P.~Antolin was partially supported by the European Research Council through the H2020 ERC Advanced Grant 2015 n.694515 CHANGE.
	G.~Sangalli was partially supported by the European Research Council through the FP7 ERC Consolidator Grant n.616563 HIGEOM.
	
	\newpage
	\bibliographystyle{unsrt}   

\end{document}